\documentclass{elsart}

\usepackage[dvips]{graphicx}
\usepackage{amsmath}

\newcommand{\bk}{{\bf k}}
\newcommand{\bq}{{\bf q}}
\newcommand{\bx}{{\bf x}}
\newcommand{\br}{{\bf r}}
\newcommand{\eF}{\varepsilon_{\mathrm{F}}}
\newcommand{\vF}{v_{\mathrm{F}}}
\newcommand{\kB}{k_{\mathrm{B}}}
\newcommand{\kF}{k_{\mathrm{F}}}
\let\onlinecite\cite

\begin{document}

\runauthor{Angilella, Leys, March, and Pucci}
\runtitle{Correlation between characteristic energies ...}
\journal{Phys. Lett. A (accepted)}
\received{\today}

\begin{frontmatter}
\title{Correlation between characteristic energies
   in non--\lowercase{$s$}-wave pairing
   superconductors}

\author[CT]{G. G. N. Angilella}
\author[RUCA]{F. E. Leys}
\author[Oxford,RUCA]{N. H. March}
\and
\author[CT]{R. Pucci}
\address[CT]{Dipartimento di Fisica e Astronomia, Universit\`a di
   Catania, and Istituto Nazionale per la Fisica della Materia, UdR
   di Catania,\\ Via S. Sofia, 64, I-95123 Catania, Italy}
\address[RUCA]{Department of Physics, University of Antwerp (RUCA),\\
   Groenenborgerlaan 171, B-2020 Antwerp, Belgium}
\address[Oxford]{Oxford University, Oxford, England}

\date{\today}

\begin{abstract}
By solution of the Bethe-Goldstone equation for the Cooper pairing
   problem, an approximate analytic relation is derived between
   coherence length $\xi$ and the binding energy of the Cooper pair.
This relation is then qualitatively confirmed by numerically
   solving the corresponding self-consistent gap equations, following
   the crossover 
   from weak to strong coupling, in non--$s$-wave superconductors.
The relation applies to non-conventional superconductors, and in
   particular to heavy Fermions and to high-$T_c$ cuprates.
Utilizing in addition a phenomenological link between $\kB T_c$ and a
   characteristic energy $\varepsilon_c = \hbar^2 / 2m^\ast \xi^2$, with
   $m^\ast$ the effective mass, major differences are exposed in the
   functional relation between $\kB T_c$ and $\varepsilon_c$ for $s$-wave
   materials and for non-conventional superconductors.
The relation between critical temperature and $\varepsilon_c$ thereby
   proposed correctly reflects the qualitative properties of heavy
   Fermion superconductors.
\begin{keyword}
\PACS
74.20.-z,
74.72.-h, 
74.70.Tx 
\end{keyword}
\end{abstract}
\end{frontmatter}

\newpage

\section{Introduction}
\label{sec:intro}

In earlier work \cite{Angilella:00b,Angilella:01a}, we have discussed
   the possible correlation between the critical temperature $T_c$ in
   anisotropic superconductors and a `natural' energy scale
   $\varepsilon_c \sim \hbar^2 / m^\ast l_c^2$ involving the effective
   mass $m^\ast$ and some characteristic length $l_c$.
Here, by anisotropic superconductors we mean a superconductor
   characterized by non-spherically symmetric pairing, giving rise to
   an anisotropic $\bk$-dependence of the gap energy in momentum
   space.
Well-known instances of such materials are the heavy Fermion
   compounds, most of them being characterized by a $p$-wave order
   parameter \cite{Sigrist:91,Heffner:95}, and the high-$T_c$,
   cuprates, whose order parameter displays $d$-wave symmetry
   \cite{Annett:90a}.
The most natural physical choice for the characteristic length $l_c$
   entering the definition of $\varepsilon_c$ above was the coherence
   length $\xi$ \cite{Angilella:00b}.
In the case of anisotropic superconducting materials in the presence
   of magnetic fluctuations, we later correlated the coherence length
   $\xi$ to the spin-fluctuation temperature $T_{\mathrm{sf}}$ as $\kB
   T_{\mathrm{sf}} \sim \hbar^2 / m^\ast \xi^2$ \cite{Angilella:01a}.

The possible enhancement of the critical temperature $T_c$ in an
   anisotropic superconductor was already studied in the early work of
   Markowitz and Kadanoff \cite{Markowitz:63} within a weak-coupling,
   BCS-like approach, and more recently revived in
   Refs.~\onlinecite{Whitmore:84,Beal-Monod:95,Valls:95} as possibly
   relevant for the high-$T_c$ cuprates.
A generalization of Markowitz and Kadanoff's results to the strong and
   intermediate coupling regime however showed that in the strong
   coupling limit anisotropy is effectively averaged out, and the gap
   tends to become isotropic \cite{Combescot:91}.

In the case of the high-$T_c$ cuprates, the strong coupling limit
   corresponds to the underdoped region of their phase diagram, which
   is usually interpreted in terms of a crossover between
   Bose-Einstein condensation of strongly coupled preformed pairs
   above $T_c$, and weak coupling, BCS-like superconductivity in the
   overdoped regime \cite{Randeria:95}.
On the other hand, the coherence length $\xi$ can be continuously
   connected to the characteristic size of the preformed bosonic pairs
   in the normal state of (underdoped) high-$T_c$ cuprates.
Therefore, the coherence length may serve to parametrize the crossover
   from weak to strong coupling, with $\kF \xi$ decreasing in going
   from weak coupling, characterized by large superconducting pair
   fluctuations, to strong coupling, $\kF$ being the Fermi momentum
   \cite{Pistolesi:94} (see also Refs.~\onlinecite{Andrenacci:00,Perali:02}, and
   refs. therein).
In the underdoped regime, a relatively short coherence
   length is consistent with the idea of preformed pairs localized in
   real space \cite{Nozieres:85,Egorov:94,March:94,Timusk:99}.
In such a limit, it is interesting to investigate how the internal
   structure of these preformed pairs is related to the overall
   symmetry of the many-body order parameter \cite{Andrenacci:03}.

Here, we derive an explicit expression for the
   coherence length in terms of the pair binding energy for the Cooper
   problem in anisotropic superconductors.
Such an expression [see Eq.~(\ref{eq:xiellneq0}) below] explicitly
   contains the quantum number $\ell$ of the pair relative angular
   momentum, which is usually employed to parametrize the anisotropic
   character of the order parameter, $\ell=0,1,2$ corresponding to
   $s$-, $p$-, and $d$-wave symmetry, respectively.
While this expression correctly reduces to the standard one for
   isotropic $s$-wave superconductors, in the case $\ell>0$ it agrees
   qualitatively with the phenomenological dependence of $\kB
   T_c$ on the characteristic energy $\varepsilon_c$, proposed in
   Ref.~\onlinecite{Angilella:00b} for the heavy Fermion compounds as
   well as for the high-$T_c$ cuprates.
Such a dependence of the characteristic energy for superconductivity
   on $\varepsilon_c$ is 
   then qualitatively confirmed by numerically solving the
   self-consistent gap equations for the maximum gap at $T=0$ in the
   case of an anisotropic superconductor in the crossover between the
   weak- and strong-coupling limits, as a function of the
   dimensionless crossover parameter $\kF \xi$, now employing a more
   general definition of the coherence length $\xi$ for anisotropic
   superconductors.

The paper is organized as follows.
In Sec.~\ref{sec:model}, after briefly reviewing the Bethe-Goldstone
   equation for the Cooper problem in the case of isotropic
   superconductors, we generalize and solve it for non--$s$-wave
   superconductors, characterized by a superconducting instability in
   the $(\ell,m)$ channel of relative angular momentum quantum
   numbers.
Our results are discussed in relation to the
   earlier phenomenological findings in
   Ref.~\onlinecite{Angilella:00b}.
In Sec.~\ref{sec:crossover} we analyze the gap equations arising from
   the corresponding many-body problem at the mean-field level.
Although we now have to resort to numerical integration (at least in
   the non--$s$-wave case), our results qualitatively confirm the
   approximate relation between characteristic energies derived in
   Sec.~\ref{sec:model}, also in the crossover from weak to
   strong coupling.
In Sec.~\ref{sec:conclusions}
   we eventually summarize and propose some directions for future work.

\section{Bethe-Goldstone equation for non--\lowercase{$s$}-wave superconductors}
\label{sec:model}

The Bethe-Goldstone equation for the Cooper
   problem \cite{Bethe:57} in momentum space reads:
\begin{equation}
(\varepsilon - 2\xi_\bk )\psi_\bk = \sum_{\bk^\prime}
   V_{\bk\bk^\prime} \psi_{\bk^\prime} ,
\label{eq:BG}
\end{equation}
where $\psi_\bk$ is the Fourier transform of the pair wave-function
   with wave-vector $\bk$ (here, we assume a spin-singlet state, with
   zero total momentum), $\xi_\bk = \hbar^2 k^2 / (2m^\ast ) - \mu$ is the
   energy of a single electron with respect to the chemical
   potential $\mu$,
   $m^\ast$ is the effective mass,
   $\varepsilon$ is the binding energy of the electron pair, and
   $V_{\bk\bk^\prime}$ is the Fourier transform of the
   electron-electron interaction (see also Ref.~\onlinecite{Ketterson:99}
   for a pedagogical review).
In the weak-coupling limit, we may safely assume $\mu=\eF$
   at $T=0$, with $\eF$ the Fermi energy.
However, we anticipate that this identification will be superseded in
   Sec.~\ref{sec:crossover}, where the crossover from weak to strong
   coupling will be addressed more in detail.
In the case of anisotropic superconductors, we adopt the spirit of the
   Anderson-Brinkman-Morel model for $p$-wave superfluidity in $^3$He
   \cite{Anderson:61}, and expand the electron-electron interaction in
   spherical harmonics around the Fermi surface as
\begin{equation}
V_{\bk\bk^\prime} = 
 -\frac{1}{\Omega} \sum_{\ell=0}^\infty \sum_{m=-\ell}^\ell V_{\ell
   m} Y_{\ell m} (\hat{\bk}) Y_{\ell m} (\hat{\bk}^\prime ),
\label{eq:Yexpansion}
\end{equation}
for $|\xi_\bk |$, $|\xi_{\bk^\prime }| < \Lambda$, and
   zero otherwise, 
   where $\hat{\bk}$ is the unit vector pointing along the direction
   of $\bk$, $\Omega$ denotes the volume of the system, and
   $\Lambda$ is an energy cut-off, characterized by the nature of
   the interaction. 
In the case of conventional superconductivity ($s$-wave pairing, or
   $\ell=0$), it would be natural to identify such an energy scale with
   the Debye energy, as in BCS theory.

The use of spherical harmonics to expand the electron-electron
   interaction in anisotropic pairing superconductors characterized by
   a spherical Fermi surface was earlier considered by Markowitz and
   Kadanoff \cite{Markowitz:63} in the weak-coupling regime (see also
   Refs.~\onlinecite{Beal-Monod:95,Valls:95}), and later by Combescot
   \cite{Combescot:91} in the strong and intermediate coupling regime.
A related model of exotic Cooper pairing with finite angular momentum
   has been discussed also in Ref.~\onlinecite{Quintanilla:01}, where
   rotational symmetry breaking ($\ell>0$) is due to the interplay
   between the finite range of the attractive potential and the
   interelectronic average distance.
Within such model, the authors of Ref.~\onlinecite{Quintanilla:01}
   also derive the $\ell$-dependence of the critical temperature in
   the Bose-Einstein limit.
In the more realistic case of a non-spherical Fermi surface, spherical
   harmonics are naturally replaced by Allen's Fermi surface harmonics
   \cite{Allen:76}, which have been used by Whitmore \emph{et al.}
   \cite{Whitmore:84} in extending the results of
   Ref.~\onlinecite{Markowitz:63} within the framework of Eliashberg
   equations for $T_c$.
Moreover, spherical harmonics afford a natural classification of
   anisotropic superconductors (such as heavy Fermion compounds
   \cite{Sigrist:91,Heffner:95} as well as high-$T_c$ cuprates
   \cite{Annett:90a}) in terms of their pairing symmetry.

Within the weak-coupling approximation, the largest attractive
   coupling constant $V_{\ell m} \equiv V > 0$ in
   Eq.~(\ref{eq:Yexpansion}) gives rise to a pairing
   instability in the $(\ell,m)$ channel.
In the following, we shall neglect other instabilities, which may
   arise well below $T_c$, corresponding to mixed symmetry pairing.
This amounts to retaining the $(\ell,m)$ term only in the expansion
   Eq.~(\ref{eq:Yexpansion}).
In this case, Eq.~(\ref{eq:BG}) has a solution given by
\begin{equation}
\psi_\bk = \alpha \frac{Y_{\ell m} (\hat{\bk})}{\varepsilon - 2 \xi_\bk}
\label{eq:BGsol}
\end{equation}
belonging to the eigenvalue
\begin{equation}
\varepsilon \simeq -2\Lambda \exp\left( - \frac{2}{VN(0)} \right),
\end{equation}
where $N(0)$ is the density of states at the Fermi level, and $\alpha$
   is a normalization constant.
Such a solution corresponds to a bound state ($\varepsilon<0$), and a
   further mean-field analysis of the many-electron problem (see also
   Sec.~\ref{sec:crossover}) shows that
   it indeed corresponds to a superconducting state, characterized by
   a gap function at $T=0$:
\begin{equation}
\Delta_{\hat{\bk}} = 2\Lambda\Gamma e^{-1/N(0)V} Y_{\ell m} (\hat{\bk}),
\end{equation}
having the symmetry of the attractive channel under consideration,
   where:
\begin{equation}
\ln\Gamma = -\int d\Omega_\bk |Y_{\ell m} (\hat{\bk})|^2 \ln |Y_{\ell m}
   (\hat{\bk})|,
\end{equation}
the integration being carried over the unit sphere \cite{March:67}.
The anisotropic $\bk$-dependence of the pair wave-function $\psi_\bk$
   in Eq.~(\ref{eq:BGsol}) in the case $\ell>0$ provides interesting
   information on the internal structure of a Cooper pair and its
   connection with the overall symmetry of the many-body gap function.

Given the pair wave-function $\psi_\bk$, the coherence length $\xi$ is
   naturally defined by
\begin{equation}
\xi^2 = \frac{\sum_\bk | \nabla_\bk \psi_\bk |^2}{\sum_\bk |\psi_\bk
   |^2} .
\label{eq:xidef}
\end{equation}
Passing to the continuous limit, in the isotropic, $s$-wave case
   ($\ell=0$), one obtains \cite{Ketterson:99}
\begin{equation}
\xi^2 = \frac{2\hbar^2}{3m^\ast \eF} \frac{1}{x^2},
\label{eq:xi0}
\end{equation}
where $x=|\varepsilon|/(2\eF)$ measures the binding energy of a pair in
   units of the energy of an unbound pair.
Apart from a numerical factor, from Eq.~(\ref{eq:xi0}) one thus recovers the
   correct order of magnitude relation between the critical
   temperature, the Fermi velocity $\vF = \hbar^{-1} d\xi_\bk /dk$,
   and the coherence length: 
\begin{equation}
\kB T_c \sim |\varepsilon| \sim \frac{\hbar \vF}{\xi} .
\label{eq:Tcxi0}
\end{equation}

In the anisotropic case, for a pairing instability in the $(\ell,m)$
   channel, with the pair wave-function given by Eq.~(\ref{eq:BGsol}),
   the denominator in Eq.~(\ref{eq:xidef}) is easily integrated in the
   continuous limit as
\begin{eqnarray}
\sum_\bk |\psi_\bk |^2  &\to& \Omega \int d^3 \bk |\psi_\bk |^2
  \nonumber \\
&&=
   \frac{\Omega \alpha^2}{4\pi} \int_0^\infty dE
   \frac{N(E)}{(\varepsilon -2E)^2} \int d\Omega_\bk |Y_{\ell m}
   (\hat{\bk})|^2 \nonumber\\
&&\approx - \Omega \alpha^2
   \frac{N(0)}{8\pi\varepsilon} ,
\label{eq:numerator}
\end{eqnarray}
where $N(E)/4\pi$ is the density of states per unit solid angle, and
   use has been made of the normalization condition of the spherical
   harmonics.
In the case $\ell\neq0$, the anisotropic $\bk$ dependence of the pair
   wave-function Eq.~(\ref{eq:BGsol}) gives rise to two contributions
   in the numerator of Eq.~(\ref{eq:xidef}), according to the chain rule
\begin{equation}
|\nabla_\bk \psi_\bk |^2 = \frac{\alpha^2}{k^2}
   \frac{|\nabla_{\hat{\bk}} Y_{\ell m}
   (\hat{\bk})|^2}{(\varepsilon-2\xi_\bk )^2} + 4 \alpha^2 \hbar^2
   \vF^2 \frac{| Y_{\ell m} (\hat{\bk})|^2}{(\varepsilon-2\xi_\bk )^4}.
\end{equation}
Here, we have made use of the decomposition $\nabla_\bk = k^{-1}
   \nabla_{\hat{\bk}} + \hat{\bk} \partial/\partial k$, where
   $\nabla_{\hat{\bk}}$ denotes the angular part of the gradient
   operator in momentum space, and of the fact that
   $\xi_\bk$ depends only on $k=|\bk|$.
Passing to the continuous limit, integrating separately over the
   angles as in Eq.~(\ref{eq:numerator}), and making use of the
   identity
\begin{eqnarray}
\int d\Omega_\bk |\nabla_{\hat{\bk}} Y_{\ell m} (\hat{\bk})|^2 &=& -
   \int d\Omega_\bk Y_{\ell m}^\ast (\hat{\bk}) \nabla_{\hat{\bk}}^2
   Y_{\ell m} (\hat{\bk}) \nonumber\\
&&= \ell(\ell+1),
\end{eqnarray}
which follows from a variant of the Green's formula over the unit
   sphere, one eventually obtains
\begin{equation}
\xi^2 = \frac{\hbar^2}{2m^\ast \eF} \left[ \frac{4}{3} \frac{1}{x^2} + \frac{\ell
   (\ell+1)}{1+x} \left( 1 - \frac{x\ln x}{1+x} \right)
   \right].
\label{eq:xiellneq0}
\end{equation}
Equation~(\ref{eq:xiellneq0}) implicitly relates the binding energy
   $|\varepsilon|$ of a Cooper pair to its characteristic size $\xi$
   for anisotropic pairing superconductors.

In the case of non--$s$-wave superconductors,
   Equation~(\ref{eq:xiellneq0}) manifestly includes an `orbital'
   contribution, proportional to $\ell(\ell+1)$, as earlier surmised
   in Ref.~\cite{Angilella:00b} on the basis of phenomenological
   considerations.
There, we proposed the existence of
   a phenomenological relation linking $\kB T_c$ to the characteristic
   energy 
\begin{equation}
\varepsilon_c = \frac{\hbar^2}{2 m^\ast \xi^2 },
\label{eq:00b}
\end{equation}
in the case of anisotropic superconductors. 
That the effective mass $m^\ast$ should enter inversely in determining
   the scale of $\kB T_c$ was earlier recognized by Uemura \emph{et
   al.} \cite{Uemura:91}, who did not, however, include the coherence
   length in their analysis \cite{Pistolesi:94}.
By comparing the experimental values for several heavy Fermion
   compounds ($p$-wave superconductors, $\ell=1$) as well as
   high-$T_c$ superconductors ($d$-wave superconductors, $\ell=2$), in
   the anisotropic case we
   found that $\kB T_c = f(\varepsilon_c)$ deviates from the square-root
   behavior, $\kB T_c \propto \sqrt{\varepsilon_c}$, that is easily
   derived from Eqs.~(\ref{eq:Tcxi0}) and (\ref{eq:00b}) for isotropic, $s$-wave
   superconductors. 
In particular, in the heavy Fermion case, we noticed a large initial
   slope in $f(\varepsilon_c )$ for $\varepsilon_c =0$, and a tendency of
   such function to approach saturation, as $\varepsilon_c$ increases.

\begin{figure}
\centering
\includegraphics[height=\columnwidth,angle=-90]{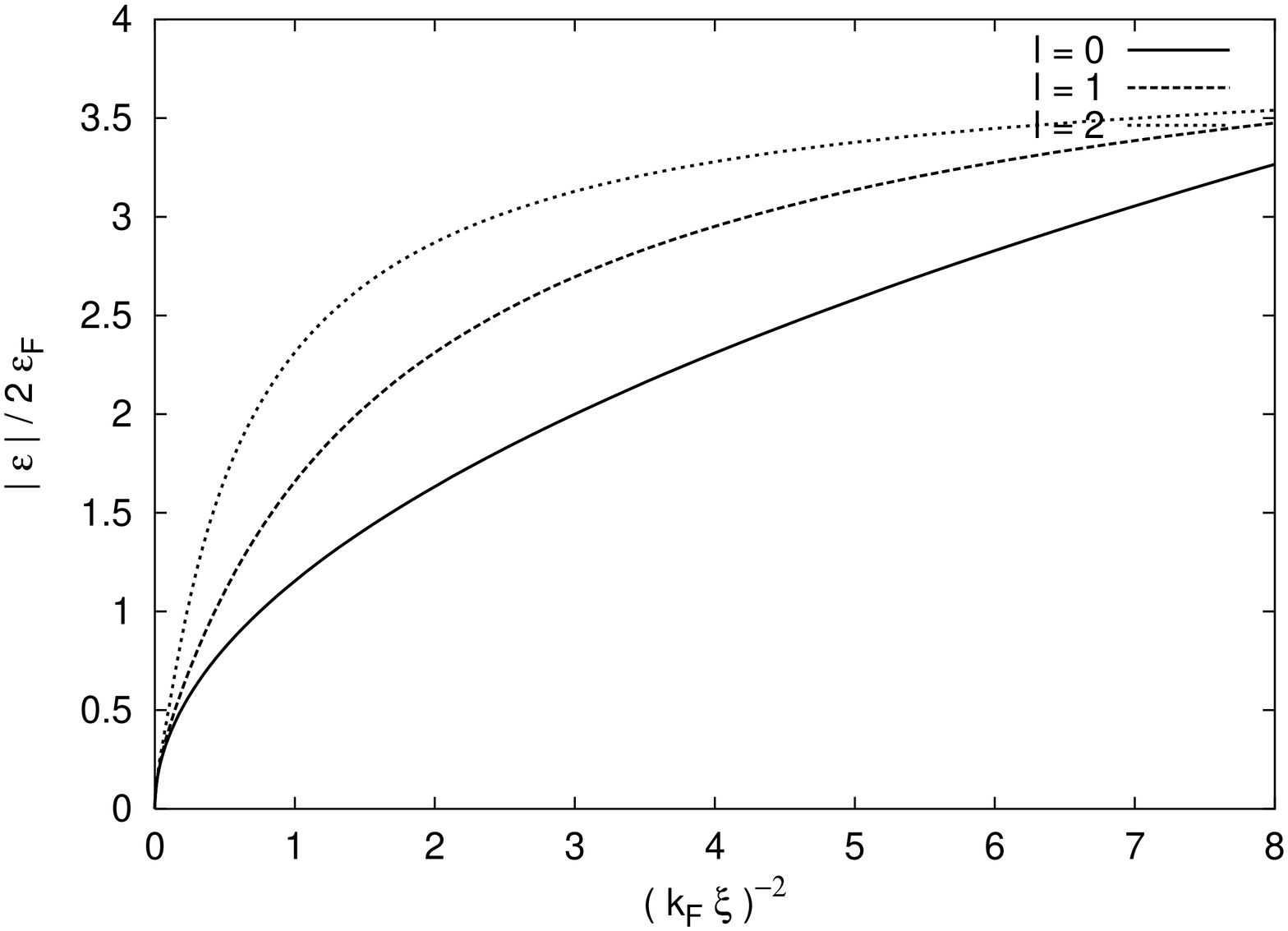}
\caption{Dependence of the normalized Cooper pair binding energy
   $|\varepsilon|/2\eF$ 
   on the characteristic energy $\varepsilon_c /\eF = [\hbar^2 /
   (2m^\ast \xi^2)]/\eF = (\kF\xi)^{-2}$, 
   as implicitly defined by Eq.~(\protect\ref{eq:xiellneq0}), for
   $\ell=0,1,2$, corresponding to $s$-, $p$-, and $d$-wave
   superconductivity, respectively.
See text for discussion.
}
\label{fig:multil}
\end{figure}

In view of the fact that $\kB T_c \sim |\varepsilon| \sim x \eF$
   \cite{Ketterson:99}, 
   Eq.~(\ref{eq:xiellneq0}) implicitly defines the generalization to
   the anisotropic case of the relationship between $\kB T_c$ and
   $\varepsilon_c$ we were looking for.
Equation~(\ref{eq:xiellneq0}) correctly reduces to Eq.~(\ref{eq:xi0}) in the
   isotropic case ($s$-wave superconductors, $\ell=0$).
In the case of non--$s$-wave superconductors ($\ell>0$),
   Eq.~(\ref{eq:xiellneq0}) indeed increases with increasing $\varepsilon_c$,
   starting with a logarithmically infinite slope at $\varepsilon_c =
   0$, and tending to a saturation value as $\varepsilon_c \to\infty$.
Such a limit is \emph{e.g.} approached for $\kF \xi\ll 1$.
Setting $a=4/[3\ell(\ell+1)]$ and performing an asymptotic analysis of
   Eq.~(\ref{eq:xiellneq0}), in the limit $\kF \xi\ll 1$ ($\ell>0$) one
   obtains
\begin{equation}
x = \frac{1+a}{W [(1+a)/e]},
\label{eq:Lambert}
\end{equation}
where $W(z)$ is Lambert's function \cite{Lambert}, and in the limit of
   very large anisotropy ($\ell\gg1$): 
\begin{equation}
\lim_{\ell\to\infty} x = W^{-1} (e^{-1} ) = 3.591\ldots .
\label{eq:Lambertinf}
\end{equation}
Figure~\ref{fig:multil} shows the dependence of the dimensionless
   measure of the
   pair binding energy $x=|\varepsilon|/2\eF$ on $\varepsilon_c / \eF
   = (\kF \xi)^{-2}$,
   implicitly defined by
   Eq.~(\ref{eq:xiellneq0}) in the cases $\ell=0,1,2$, corresponding
   to $s$-, $p$-, and $d$-wave superconductivity, respectively.
Keeping fixed all other variables, one notices that for $\xi>\xi_0$,
   where
\begin{equation}
\frac{\hbar^2}{2m^\ast \xi_0^2 \eF} = \frac{3}{4W^2 (e^{-1})} = 9.672\ldots,
\label{eq:Lambertxi0}
\end{equation} 
$|\varepsilon|$ increases as $\ell$ increases while,
   for $\xi<\xi_0$, $|\varepsilon|$ decreases as $\ell$ increases, although
   without large deviations from the limiting value,
   Eq.~(\ref{eq:Lambertinf}).
Therefore, at least for sufficiently weakly coupled superconductors
   ($\xi>\xi_0$), anisotropy enhances superconductivity, the 
   degree of anisotropy being here parametrized by the order $\ell$ of
   the spherical harmonic modelling the $\bk$ dependence of the order
   parameter.

\section{Mean-field analysis of the many-body problem and the
   crossover from weak to strong coupling}
\label{sec:crossover}

The definition of the Cooper pair size $\xi$ we employ in
   Eq.~(\ref{eq:xidef}) makes use of the pair wave-function $\psi_\bk$
   for the Cooper problem, in which the only many-body effect is that
   of forbidding the occupancy of states below the Fermi level, $\eF$.
Within BCS theory, \emph{i.e.} at the mean field level, a
   self-consistent treatment of the superconducting instability
   affords a better definition of the pair wave-function, which in
   momentum space is given by \cite{Ketterson:99} $\psi_\bk =
   \Delta_\bk /(2 E_\bk )$, where $\Delta_\bk$ is the gap function, now
   depending on $\bk$ as a vector, and $E_\bk = (\xi_\bk^2 +
   |\Delta_\bk |^2 )^{1/2}$ the upper branch of the quasiparticle
   dispersion relation.
With this definition of $\psi_\bk$, when $E_\bk$ is allowed to vanish
   locally on the Fermi surface, as is the case for $p$- and $d$-wave
   superconductors, it has been recently shown that nodal
   quasiparticles give rise to a logarithmically divergent
   contribution to $\xi$, as defined by Eq.~(\ref{eq:xidef})
   \cite{Benfatto:02}.
While such a drawback does not arise in $s$-wave superconductors
   \cite{Marini:98}, in the case of anisotropic superconductors the
   coherence length must be defined in terms of the range
   in real space of
   the static correlation function for the modulus of the
   order parameter \cite{Pistolesi:96,Benfatto:02}.
The fact that our result Eq.~(\ref{eq:xiellneq0}) provides a finite
   estimate for $\xi$ also in the anisotropic case clearly depends on
   our approximate choice for $\psi_\bk$, which solves only the
   two-body Cooper problem.

In order to
   discuss the limits 
   of validity of the approximate results 
   derived in Sec.~\ref{sec:model}, we will then work out numerically
   the mean-field solution to the corresponding BCS Hamiltonian, both
   in the weak and in the strong-coupling limit.
We start by reviewing the model and notations set out in
   Refs.~\cite{Pistolesi:94,Pistolesi:96,Marini:98,Benfatto:02} (see also
   Ref.~\cite{Babaev:01}).

We shall consider the following Hamiltonian for a superconducting
   system in three dimensions \cite{note:static}:
\begin{equation}
H = \sum_{\bk\sigma} \xi_\bk c^\dag_{\bk\sigma} c_{\bk\sigma} +
   \sum_{\bk\bk^\prime \bq} V_{\bk\bk^\prime} c^\dag_{\bk\uparrow}
   c^\dag_{-\bk+\bq\downarrow} c_{-\bk^\prime+\bq\downarrow}
   c_{\bk^\prime\uparrow} , 
\label{eq:hamiltonian}
\end{equation}
where $c^\dag_{\bk\sigma}$ [$c_{\bk\sigma}$] is the creation
   [destruction] operator for an electron with wave-vector $\bk$ and
   spin projection $\sigma\in\{\uparrow,\downarrow\}$ along a
   specified direction, $\xi_\bk = \hbar^2 k^2 /(2m^\ast ) -\mu$ is
   the dispersion relation for free electrons with effective mass
   $m^\ast$, measured with respect to the chemical potential $\mu$,
   and
\begin{equation}
V_{\bk\bk^\prime } = -\frac{V}{\Omega}
 Y_{\ell m} (\hat{\bk}) Y_{\ell m} (\hat{\bk}^\prime )
\label{eq:interY}
\end{equation}
is the projection of the electron-electron interaction along the
   $(\ell,m)$ channel, as in
   Eq.~(\ref{eq:Yexpansion}), which we assume
   to be attractive ($V>0$) \cite{note:Coulomb}.
Equation~(\ref{eq:interY}) generalizes to the non--$s$-wave case the
   contact potential in real space discussed \emph{e.g.} by Marini
   \emph{et al.} \cite{Marini:98}.
Standard diagonalization techniques then lead to the mean-field
   coupled equations for the gap energy $\Delta_\bk = \Delta_0 Y_{\ell
   m} (\hat{\bk})$ and the particle density $n$ at $T=0$ (see
   Ref.~\cite{Babaev:01} for the case $T\neq0$):
\begin{subequations}
\begin{eqnarray}
\label{eq:BCSa}
\frac{1}{V} &=& \frac{1}{\Omega} \sum_\bk \frac{|Y_{\ell m}
   (\hat{\bk})|^2}{2E_\bk} ,\\ 
\label{eq:BCSb}
n &=& \frac{2}{\Omega} \sum_\bk v^2_\bk ,
\end{eqnarray}
\label{eq:BCS}
\end{subequations}
where $v_\bk^2 = \frac{1}{2} (1 - \xi_\bk /E_\bk )$, together with
   $u_\bk^2 = 1-v_\bk^2$, are the usual coherence factors of BCS
   theory.

Owing to our choice of a free particle dispersion relation (all band
   effects are embedded in a single parameter, namely the effective
   mass $m^\ast$), and of a contact potential in real space, the sums
   over $\bk$ in Eq.~(\ref{eq:BCSa}) above 
   give rise to an ultraviolet divergence, which requires a suitable
   regularization.
In three dimensions, it is customary to intoduce the scattering
   amplitude $a_s$ \cite{Pistolesi:96,Marini:98,Pieri:00}, which for our
   anisotropic interaction reads:
\begin{equation}
\frac{m^\ast}{4\pi a_s} = -\frac{1}{V} + \frac{1}{\Omega} \sum_\bk
   \frac{m^\ast}{k^2} |Y_{\ell m} (\hat{\bk})|^2 .
\label{eq:scatt}
\end{equation}
Subtracting Eq.~(\ref{eq:scatt}) from Eq.~(\ref{eq:BCSa}) one has:
\begin{equation}
-\frac{m^\ast}{4\pi a_s} = \frac{1}{\Omega} \sum_\bk \left(
   \frac{1}{2E_\bk} - \frac{m^\ast}{k^2} \right) |Y_{\ell m}
   (\hat{\bk})|^2 .
\label{eq:BCSc}
\end{equation}
Following Ref.~\cite{Marini:98}, we render Eqs.~(\ref{eq:BCS}) and
   Eq.~(\ref{eq:BCSc}) dimensionless, by introducing the dimensionless
   quantities 
\begin{alignat*}{2}
x^2 &= \frac{\hbar^2 k^2}{2m^\ast} \frac{1}{\Delta_0} ,
& \quad x_0 &= \frac{\mu}{\Delta_0} ,\\
\xi_x &= \frac{\xi_\bk}{\Delta_0} = x^2 -x_0 , & \quad E_x &=
   \frac{E_\bk}{\Delta_0} = \sqrt{\xi_x^2 + |Y_{\ell m} (\hat{\bx})|^2}
   ,
\end{alignat*}
and the Fermi energy $\eF = \hbar^2 \kF^2 /(2m^\ast ) = \hbar^2
   (3\pi^2 n)^{2/3} /(2m^\ast )$.
In particular, Marini \emph{et al.} \cite{Marini:98} observe that $x_0
   =\mu/\Delta_0$ can be used as a parameter for the crossover between
   the strong-coupling, Bose-Einstein (BE) limit ($x_0 \ll 0$) and the
   weak-coupling, BCS limit ($x_0 \gg 0$).
Moreover, on one hand, the expression resulting from the dimensionless version
   of Eq.~(\ref{eq:BCSc}) for $(\kF a_s )^{-1}$ can be inverted to
   obtain $x_0$ as a function of the dimensionless scattering length
   $\kF a_s$.
On the other hand, one may alternatively use $\kF \xi$ as the
   independent variable in place of $\kF a_s$ \cite{Marini:98}.
Indeed, it was earlier recognized by Pistolesi and Strinati
   \cite{Pistolesi:94} (following the seminal work of Nozi\`eres and
   Schmitt-Rink \cite{Nozieres:85}) that a natural variable which can
   be used to follow the crossover from strong-coupling to weak-coupling
   superconductivity is the product $\kF\xi$ of Fermi wave-vector
   $\kF$ times the coherence length $\xi$ for two-electron
   correlation.
In the BE limit, electrons are expected to bind in quasi-bound pairs
   localized in real space (Schafroth pairs \cite{Schafroth:54}), thus
   realizing the condition $\kF \xi \ll 1$, while the BCS limit
   corresponds to loosely coupled pairs, with $\kF \xi \gg 1$,
   localized in momentum space close to the Fermi energy.

A zero-temperature calculation of the coherence length for a
   three-dimensional, $s$-wave 
   superconductor along the crossover between the weak- and the
   strong-coupling limits has been performed both numerically
   \cite{Pistolesi:96} and analitically \cite{Marini:98}.
The case of a two-dimensional, $d$-wave superconductor has been
   discussed in Ref.~\cite{Benfatto:02}, where a dispersion relation
   typical of the cuprate superconductors has been explicitly
   considered.
As anticipated above, in the non--$s$-wave case ($\ell\neq0$), the
   standard definition, Eq.~(\ref{eq:xidef}), of the coherence length
   leads to unphysical divergences.
One may still conveniently define a coherence length as the range in
   real space of the static correlation function $X_\Delta (\br)$ for
   the modulus of the order parameter \cite{Pistolesi:96,Benfatto:02}
   as
\begin{equation}
\xi^{-1} = - \lim_{r\to\infty} \frac{\log X_\Delta (r)}{r} .
\end{equation}
The Fourier transform $X_\Delta (\bq)$ of such a function has been
   derived for an $s$- and a $d$-wave superconductor in
   Refs.~\cite{Pistolesi:96} and \cite{Benfatto:02}, respectively.
In the case of our anisotropic interaction, Eq.~(\ref{eq:interY}), it
   reads:
\begin{eqnarray}
X_\Delta (\bq)^{-1} &=& \frac{1}{V} - \frac{1}{2\Omega} \sum_\bk 
   |Y_{\ell m} (\hat{\bk})|^2 \frac{(u_{\bk +\bq/2} u_{\bk-\bq/2} -
   v_{\bk+\bq/2} v_{\bk-\bq/2} )^2}{E_{\bk+\bq/2} + E_{\bk-\bq/2}}
   \nonumber \\
&=& \frac{1}{2\Omega} \sum_\bk |Y|^2 \left[
   \frac{1}{E} - \frac{1}{E_+ + E_-} \left( 1 + \frac{\xi_+ \xi_- -
   |\Delta_+ | |\Delta_- |}{E_+ E_- } \right) \right],
\label{eq:LaraX}
\end{eqnarray}
where $E\equiv E_\bk$, $Y\equiv Y_{\ell m} (\hat{\bk})$, $\xi_\pm$,
   $\Delta_\pm$, 
   $E_\pm$ are calculated at momenta $\bk \pm \bq/2$, respectively,
   and use has been made of the gap equation, Eq.~(\ref{eq:BCSa}), in
   going into the second line.

First of all, since the summand in Eq.~(\ref{eq:LaraX}) depends only
   on $k=|\bk|$, $q=|\bq|$, and on the relative angle between $\bk$
   and $\bq$, passing to the continuum limit and transforming the sums
   over wave-vectors into an integral, one has $X_\Delta (\bq )\equiv
   X_\Delta (q)$.
Moreover, $X_\Delta (q)$ is an even function of its argument.
Back to real space, one analogously has $X_\Delta (\br )\equiv
   X_\Delta (r)$.
Then, the asymptotic behaviour of $X_\Delta (r)$ at large distance $r$
   will be governed by the behaviour of its Fourier transform
   $X_\Delta (q)$ at small wave-vector $q$.
In particular, assuming the expansion \cite{Pistolesi:96,Benfatto:02}:
\begin{equation}
X_\Delta^{-1} (q) = a + b q^2 + O(q^4 ),
\end{equation}
it is straightfoward to show \cite{Pistolesi:96} that
\begin{equation}
\xi^2 = \frac{b}{a} .
\label{eq:xidefnew}
\end{equation}
Such a definition of the coherence length is now consistent also for
   nodal superconductors \cite{Benfatto:02}.
It reduces to Eq.~(\ref{eq:xidef}) in the weak-coupling limit for the
   $s$-wave case, and applies also to the strong-coupling regime,
   provided $b>0$ (given that $a>0$, identically, provided $\Delta_0
   \neq0$), \emph{i.e.} 
   provided that $X_\Delta (q)$ has its absolute minimum at $q=0$
   \cite{note:Lara}.

\begin{figure}
\centering
\includegraphics[height=\columnwidth,angle=-90]{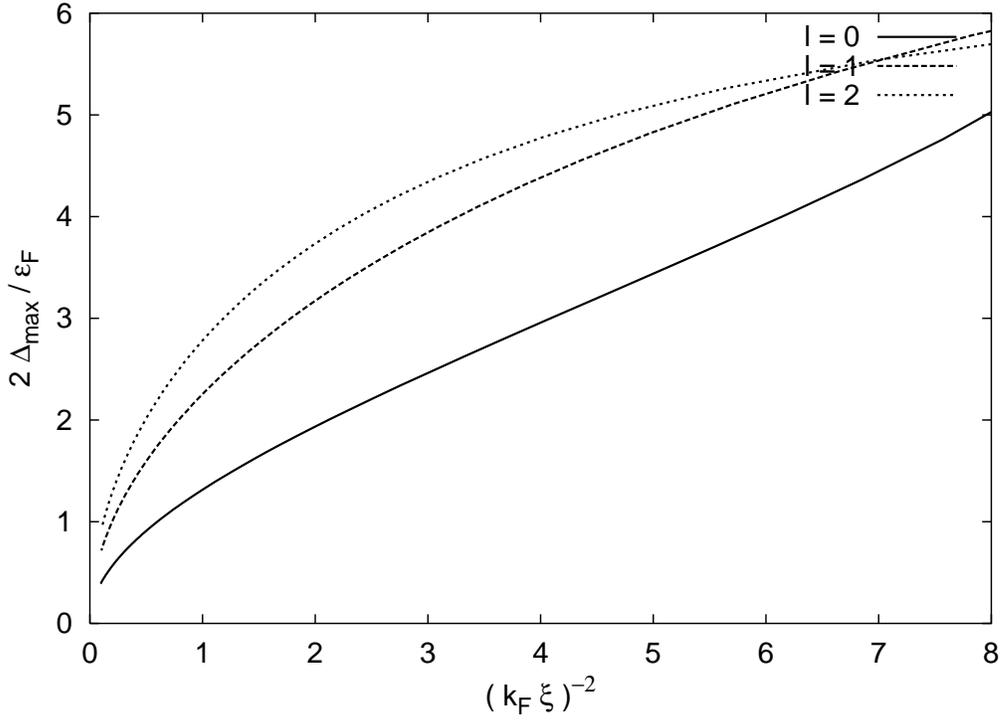}
\caption{Dependence of twice the normalized maximum gap energy
   $2\Delta_{\mathrm{max}}/\eF$ 
   on the characteristic energy $\varepsilon_c /\eF = (\kF \xi)^{-2}$, for
   $\ell=0,1,2$, corresponding to $s$-, $p$-, and $d$-wave
   superconductivity, respectively.
As $\kF\xi$ decreases [\emph{i.e.,} $(\kF\xi)^{-2}$ increases] one
   crosses over from the weak-coupling, BCS limit, to the
   strong-coupling, BE limit.
}
\label{fig:delta}
\end{figure}

We have numerically solved the gap equations, Eqs.~(\ref{eq:BCS}), and
   evaluated the dimensionless coherence length $\kF\xi$ according to
   this more general definition, Eq.~(\ref{eq:xidefnew}), for the
   anisotropic potential Eq.~(\ref{eq:interY}), with $\ell=0,1,2$,
   corresponding to $s$-, $p$-, and $d$-wave pairing, respectively.
Here, a convenient measure of the characteristic energy for
   superconductivity may be taken as the maximum gap
   $\Delta_{\mathrm{max}}$, where
   $\Delta_{\mathrm{max}} \propto \kB T_c$ holds also for anisotropic
   superconductors \cite{Combescot:91}.
Such a quantity is readily extracted
   from the solution of the gap equations, Eqs.~(\ref{eq:BCS}).
In Figure~\ref{fig:delta}, we plot the dimensionless characteristic
   energy $2\Delta_{\mathrm{max}}/\eF$ versus $\varepsilon_c /\eF =
   (\kF \xi)^{-2}$ in the cases $\ell=0,1,2$.
As $\kF \xi$ decreases [$(\kF \xi)^{-2}$ increases], one crosses over
   from the weak-coupling, BCS limit into the strong-coupling, BE
   limit \cite{Pistolesi:94}.
Despite a mean-field solution of the self-consistent gap equations has
   been now taken into account in the calculation, and a more general
   and consistent definition of the coherence length has been
   employed, our numerical results are in good qualitative agreement
   with the approximate results of Sec.~\ref{sec:model}, with
   $2\Delta_{\mathrm{max}}/\eF$ increasing as a function of
   $\varepsilon_c /\eF$ with a steeper slope, as $\ell$ increases from
   $\ell=0$ ($s$-wave) to $\ell=2$ ($d$-wave), all curves tending to
   saturation, at least within the numerically accessible range of the
   crossover parameter $\kF \xi$.
These results are in agreement with the phenomenological plots correlating
   characteristic energies for both the heavy Fermion and cuprate
   superconductors in Refs.~\cite{Angilella:00b,Angilella:01a}.

\section{Summary and future directions}
\label{sec:conclusions}

We have studied the Cooper problem for an anisotropic superconductor
   characterized by an electron-electron interaction expanded in terms
   of simple spherical harmonics over the Fermi sphere.
In the weak-coupling limit of a superconducting instability in the
   $(\ell,m)$ channel, we have derived an analytical expression for
   the relation between the pair binding energy and the correlation
   length, for arbitrary relative angular momentum quantum number
   $\ell$.
While such an expression correctly reduces to the standard one in the
   $s$-wave case ($\ell=0$), in view of the fact that $\kB T_c$ scales
   with such a pair binding energy, in the case of non--$s$-wave
   superconductors ($\ell>0$) our expression agrees qualitatively
   with the phenomenological correlation between $\kB T_c$ and
   the characteristic energy $\hbar^2 / (2 m^\ast \xi^2 )$ earlier found in
   Ref.~\onlinecite{Angilella:00b}.
These results have been confirmed by a numerical solution of the
   self-consistent gap equations in the crossover between the weak-
   and the
   strong-coupling limits, where a more general definition for the
   coherence length has been employed, applying to anisotropic
   superconductors.

It may, in the future, prove of significance to include the effect of
   Coulomb interaction on the formation of Cooper pairs in
   superconducting assemblies along lines such as those laid down in
   Ref.~\onlinecite{Lal:92b} (see also Ref.~\onlinecite{Mila:91}).
The question arises then as to whether, in the future, it will be of
   significance in making the present study the basis of fully
   quantitative calculations, to generalize beyond the result here,
   and of course, of many earlier workers, that the Cooper pairs
   formed as a consequence of an attractive interaction correspond to
   a single value of the binding energy at a given temperature.
There is some evidence (see, \emph{e.g.,} Refs.~\onlinecite{Lal:91,Lal:92a})
   that some physical properties, and in particular specific heat and
   tunneling spectra of cuprate materials, may require generalization
   to Cooper pairs with a finite range of binding energies.
We do not anticipate that the qualitative trends proposed in the
   present paper will be grossly affected.

\begin{ack}
G.G.N.A. gratefully acknowledges Dr. N. Andrenacci for useful and
   stimulating discussions.
F.E.L. thanks Professor K. Van Alsenoy for much motivation and
   encouragement.
F.E.L. also acknowledges financial support from the ``Program for the
   Stimulation of Participation in Research Programs of the European
   Union'' of the University of Antwerp.
N.H.M. and F.E.L. thank the Department of Physics and Astronomy,
   University of Catania, for the stimulating environment and much
   hospitality.
\end{ack}

\bibliographystyle{elsart-num}
\bibliography{a,b,c,d,e,f,g,h,i,j,k,l,m,n,o,p,q,r,s,t,u,v,w,x,y,z,zzproceedings,Angilella,notes}

\begin{thebibliography}{10}
\bibitem{note:Coulomb}
Within this model, the effect of a repulsive Coulomb interaction can be
  included as an on-site repulsive Hubbard term, so that the interaction
  potential, Eq.~(\ref{eq:interY}), would read $\Omega V_{\bk\bk^\prime} = U -
  V Y_{\ell m} (\hat{\bk}) Y_{\ell m} (\hat{\bk}^\prime )$, with $U>0$. The
  character of such a contribution would then be $s$-wave, and for a separable
  choice of the interaction potential it would then affect other $s$-wave
  components only. Indeed, by numerically solving the self-consistent gap
  equations for a separable potential allowing for mixed symmetry, some of the
  present authors have shown that a repulsive onsite $U>0$ would reduce the
  mean-field solution for $T_c$ associated to a superconducting instability in
  the $s$-wave channel, while leaving an instability in other, orthogonal
  channels unaffected \protect\cite{Angilella:96}. In the present case, since
  we are interested only in the superconducting instability in one symmetry
  channel at a time, the effect of a nonzero (positive) $U$ would then only
  affect the $\ell=0$ case, by a simple renormalization of the coupling
  constant ($-V_{00}/4\pi \mapsto U - V_{00}/4\pi$).
\end{thebibliography}

\newpage

\end{document}